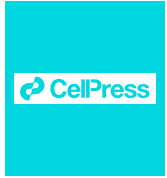
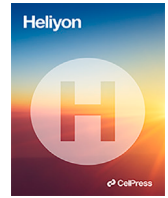

Research article

# RUL forecasting for wind turbine predictive maintenance based on deep learning

Syed Shazaib Shah [a],[*], Tan Daoliang [a], Sah Chandan Kumar [b]

[a] *School of Energy and Power, Beihang University, Beijing, 100191, PR China*
[b] *School of Software, Beihang University, Beijing, 100191, PR China*

**A B S T R A C T**

Predictive maintenance (PdM) is increasingly pursued to reduce wind farm operation and maintenance costs by accurately predicting the remaining useful life (RUL) and strategically scheduling maintenance. However, the remoteness of wind farms often renders current methodologies ineffective, as they fail to provide a sufficiently reliable advance time window for maintenance planning, limiting PdM's practicality. This study introduces a novel deep learning (DL) methodology for future-RUL forecasting. By employing a multi-parametric attention-based DL approach that bypasses feature engineering, thereby minimizing the risk of human error, two models—ForeNet-2d and ForeNet-3d—are proposed. These models successfully forecast the RUL for seven multifaceted wind turbine (WT) failures with a 2-week forecast window. The most precise forecast deviated by only 10 minutes from the actual RUL, while the least accurate prediction deviated by 1.8 days, with most predictions being off by only a few hours. This methodology offers a substantial time frame to access remote WTs and perform necessary maintenance, thereby enabling the practical implementation of PdM.

## 1. Introduction

Wind energy as the leading source of renewables has been in the lime-, driven by the heightened concerns over the environmental impacts and unsustainability of fossil-fuels for the future [1,2]. In order to keep up with the rising demand, the wind industry is actively working to make it more viable and competitive, which means tackling some of the biggest challenges it faces [3,4]. A survey analysis [5] shows that approximately 45% of the overall budget might be set aside for operation and maintenance (O&M), as shown in Fig. 1, posing as one of the biggest challenges faced by the wind industry. To counter this, preventive maintenance (PM) could be employed, which follows a periodically scheduled maintenance plan to reduce unplanned maintenance. However, this leads to unnecessary downtime, as often times the maintenance is not required [6–8]. This could be resolved if predictive maintenance (PdM) could be achieved. PdM predicts the optimal time for maintenance, ensuring it is performed precisely when needed and avoiding unnecessary machine stoppages [9]. One way to achieve this is by analyzing the remaining useful life (RUL) of the turbine and scheduling maintenance immediately prior to failure [10]. However, wind farms are often located in remote locations, usually spanning over many miles [11,12]; especially in the case of off-shore wind farms [13,14]; and timely arrival becomes an issue as a result. Consequently, current RUL methodologies fail to provide a significant reliable time window in advance for the maintenance teams to reach the site and implement an effective response [15,16]. Therefore, there is an urgent need for a methodology that can reliably forecast RUL with a significantly extended lead time.

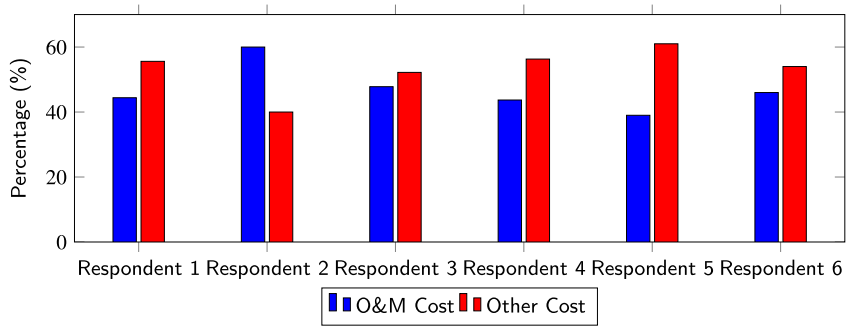

**Fig. 1.** Wind farms O&M Cost and Other Cost Breakdown from survey respondents [5].

Guo et al. [17] made use of a digital twin model to calculate RUL, while the work by Pagitsch et al. [18] modelled internal loads for predicting RUL in wind turbine (WT) gearbox. These approaches, while effective, require adequate computing power, resulting in a hindrance to industrial adoption [9]. Si et al. [19] gave a review on statistical learning-based RUL estimation that emphasizes the need to improve flexibility, adaptability, feature engineering, data utilization, and representation capability. On the other hand, Rivas et al. [20] summarized important PdM strategies and instigated the use of deep neural networks (DNN) as a cost-effective, reliable way to complete PdM tasks. Building upon this foundation, the current study also employs DNN to undertake the computation task associated with RUL. On this basis, the most common way to measure WT health degradation (which can be used to make RUL estimates) is to use one or a few predetermined parameters that show how failures progress [21,22]; known as "feature engineering." For example, Ng and Liam monitored and processed gearbox oil sump temperature for WT failure detection using kNN, XGBoost, and ANN [23]. But this approach limits the prediction capabilities to a specific failure type and a specific component. Our study did a detailed analysis of the correlation among different failure locations and found analogous behaviour between them prior to failure. Furthermore, by employing DNN to read these correlations, the RUL estimation for multifaceted failures in WT was achievable; by-passing feature engineering. To compensate for feature engineering, attention mechanism (AM) was introduced in two proposed custom models, ForeNet-2d and ForeNet-3d.

The contributions of this work are fourfold:

1. Proposing a novel concept of $RUL_f$ as a cost-effective method for forecasting upcoming WT failures, this study showcased a 2-week advance forecast. This extended forecasting window provides ample time for effective maintenance planning and supports the practical implementation of PdM for wind turbines (WTs) in its true sense.
2. Employing an analysis-based methodology to explore the interrelation among various failure types in WTs and harnessing DL to leverage this correlation, the study successfully demonstrated accurate prognostication of $RUL_f$ for a diverse range of WT failure types.
3. To achieve diversity for $RUL_f$, feature engineering was bypassed, which consequently led to improved practicality by eliminating the scope of human error during feature engineering.
4. Two custom-optimized hybrid DNN-attention-based models, ForeNet-2d (2D model architecture) and ForeNet-3d (3D model architecture), were presented for this application. Both models achieved much superior performance as compared to traditional and other famous model architectures.

The remaining portion of this paper follows: Data Procurement & Analysis, Experiment Setting, Results and Analysis, Conclusion, and Future work.

## 2. Data procurement & analysis

### 2.1. SCADA data

Currently, a cohort of research has illustrated the use of the supervisory control and data acquisition (SCADA) dataset to detect degradation in WT and its components [24,25]. DNNs can be trained to recognize these degradation patterns, allowing them to forecast WT failures. Kevin et al. [26] utilized this SCADA data and alarm history to assess WT health, allowing them to diagnose defects with 71% accuracy within a 30-hr time window. SCADA data refers to the data collected in real time from sensors, devices, and control systems located in an industrial or infrastructural environment, equipment, or machine [27]. SCADA data have shown predictive capabilities despite being initially meant for control applications only [28,29]. Due to the prevalence of SCADA systems in the WT industry, various research studies have made use of it to evaluate key operation aspects like change detection [30], fault detection [31,32], power curves [33–36], WT health [37], and more. Hu et al. [38] predicted the WT bearing RUL by using the Weiner method on SCADA data from WT bearing temperature sensors, making the case for $RUL_f$ to be based on SCADA data.

**Table 1**
WT Failure instances tally.

| Turbine Tags | Number of failures from January 1st, 2017 till December 31st, 2017 |
| --- | --- |
| T01 | 1 |
| T06 | 2 |
| T07 | 3 |
| T11 | 2 |

The dataset utilized in this research paper is an openly accessible dataset obtained from EDP (Energias de Portugal), a significant entity in the wind power industry, particularly in the Iberian Peninsula, where it holds a substantial market share in electricity production and distribution [39]. The dataset comprises SCADA data logs for a duration of one year in 2017, specifically from January 1st to December 31st. The EDP dataset contains SCADA data from four turbines, with a total of 82 parameters captured at a frequency of 10 minutes. It is widely regarded as one of the most comprehensive open-sourced databases available. While the EDP asserts that the dataset comprises logs from five turbines, it should be noted that analysis of the SCADA data showed that one of the turbine logs was absent. A detailed review by Menezes et al. [40] of this dataset provides ample information about these WTs, specifically: anemometer sensors 1 and 2 are positioned at heights of 80 m and 77 m, respectively; weather vanes are situated at heights of 77 m and 40 m; and temperature and pressure sensors are located at heights of 75 m and 100 m, respectively. A summary of the available sensor parameters in the dataset is presented as follows:

**WT Characteristics**: This section includes information on the WT's power, rotor, gearbox, generator, tower, and power curve. It provides details about the rated power, cut-in wind speed, rated wind speed, and cut-out wind speed, among others.

**Wind Speed and Direction**: This category consists of data from two anemometer sensors that measure wind speed and direction. Additionally, there are sensors for ambient temperature, air pressure, humidity, and precipitation on the meteorological mast.

**Generator RPM and Temp**: This section includes data on generator RPM (rotations per minute) and temperature.

**Component Signals**: The dataset offers information on a variety of components, such as the temperature of the gearbox oil, the temperature of the nacelle, the total active and reactive power, and the pitch angle.

**Failure Logs**: The failure logbook contains records of all WT components, documenting any failures, replacements, or repairs that have occurred. The failure logs are not part of the SCADA data as they are not sensor readings and thus are provided in a separate log file; further explanation is in Section 2.2.

*2.2. Failure logs data*

The primary rationale behind selecting EDP dataset was the inclusion of failure logs; provided in a separate log file. The initial step in data processing involved the integration of failure logs with precise time logs into the SCADA data. Upon initial analysis, it was concluded that a cumulative count of 12 failure instances occurred within the specified time period, but unfortunately, due to the lack of the corresponding SCADA data for one of the turbines and other irregularities in the dataset, only eight failure instances could be used for this application, which were separated into individual datasets and labelled with a unique "Failure Tag" in numbered fashion accordingly; details of the selected are listed in Tables 1 and 2. Referring to Table 2, Failure 6 exhibits a limited number of 196 log entries; each log entry is recorded at a 10-minute interval, resulting in a data span of approximately one and a half days only. However, as emphasized previously, the objective of this research study is to provide a significant time window for WT PdM. Hence, using Failure 6 is impractical for our particular application. After omitting Failure 6, we are left with a total of seven instances of failure, each organized into distinct datasets. The RUL forecasting was conducted using a set of seven failures, designated as distinct datasets named Failure 1, Failure 2, Failure 3, Failure 4, Failure 5, Failure 7, and Failure 8.

*2.3. Data analysis*

A thorough data analysis on all datasets was done in order to study the behaviour of WT breakdowns. A heat correlation map, amounting to all of the available parameters, is presented in Fig. 2, with sub-Fig. 2a to 2h showcasing the parametric coherence for the given datasets from Failure 1 to 8 respectively. This map highlights the correlation between the diverse parameter sets; the visual aesthetic indicates a strong coherence between multiple parameters. Under this premise, it can be postulated that, for a specific failure instance, multiple contributing variables would exhibit an unhealthy status pre-breakdown. We included the heat map for Failure 6 in Fig. 2 to examine the erratic behaviour of the WT during this instance; however, it is important to clarify that this inclusion serves solely as to further testify towards it's exclusion, as the decision to opt out Failure 6 was based on its operational duration, as outlined in Table 2. Based on the heat map, Failure 6 has a notably feeble and inconsistent pattern, which might be related to the limited amount of data log ($N$) available. The brief operational run of WT in case of Failure 6 also points towards the possibility that the WT was in aberrant status from the start-up, leading to an imminent failure, and further testified by the heat map.

The covariation between different parameter combinations is shown in Fig. 3 to examine the relation among the failure points. The data points from the last one hour prior to failure are highlighted in the plots. Two behavioural correlation clusters are observed:

- Group 1: Failure 2, Failure 4, Failure 6, Failure 7, and Failure 8

**Table 2**
Failure instances detail.

| Turbine Tags | Failure Location | Failure Detail | Failure Tag | N | Validity for $RUL_f$ |
|---|---|---|---|---|---|
| T01 | Transformer | Transformer fan damaged | Failure 1 | 31779 | Yes |
| T06 | Hydraulic group | Oil leakage in Hub | Failure 2 | 33005 | Yes |
| T06 | Gearbox | Gearbox bearings damaged | Failure 3 | 8480 | Yes |
| T07 | Hydraulic group | Oil leakage in Hub | Failure 4 | 24101 | Yes |
| T07 | Generator bearing | Generator bearings damaged | Failure 5 | 9011 | Yes |
| T07 | Generator | Generator damaged | Failure 6 | 196 | No: Too little data entries, invalid for forecasting |
| T11 | Hydraulic group | Hydraulic group error in the brake circuit | Failure 7 | 16661 | Yes |
| T11 | Hydraulic group | Hydraulic group error in the brake circuit | Failure 8 | 19814 | Yes |

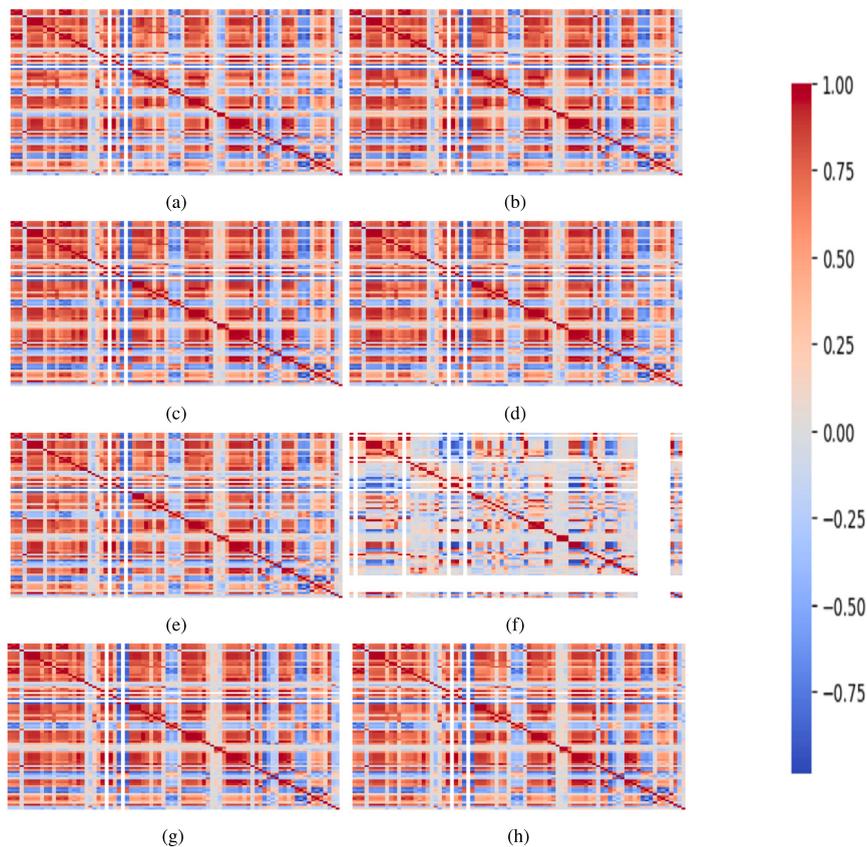

**Fig. 2.** Heat correlation map of datasets (a) Failure 1 (b) Failure 2 (C) Failure 3 (d) Failure 4 (e) Failure 5 (f) Failure 6 (g) Failure 7 (h) Failure 8.

- Group 2: Failure 1, Failure 3, Failure 5

Group 1 is associated with the hydraulic group (see Table 2), except for Failure 6, which is an outlier among the available datasets. Group 2 consists of the remaining failure datasets, each of which is linked to distinct failure sites. This interrelationship among different failure point behaviors is of key importance to allowing the DNN models to achieve robust prediction capabilities for multifaceted failure forecasting. An emphasized case on the relation between WT reactive power and rotor RPM was made by Menezes et al. [40], underscoring its role towards failure behaviour in WT, based on which a similar analysis was performed for this study; see Fig. 4. Again, similar behaviour among the grouped failure points is observed, whereas abnormal behaviour for Failure 6 can also be noticed.

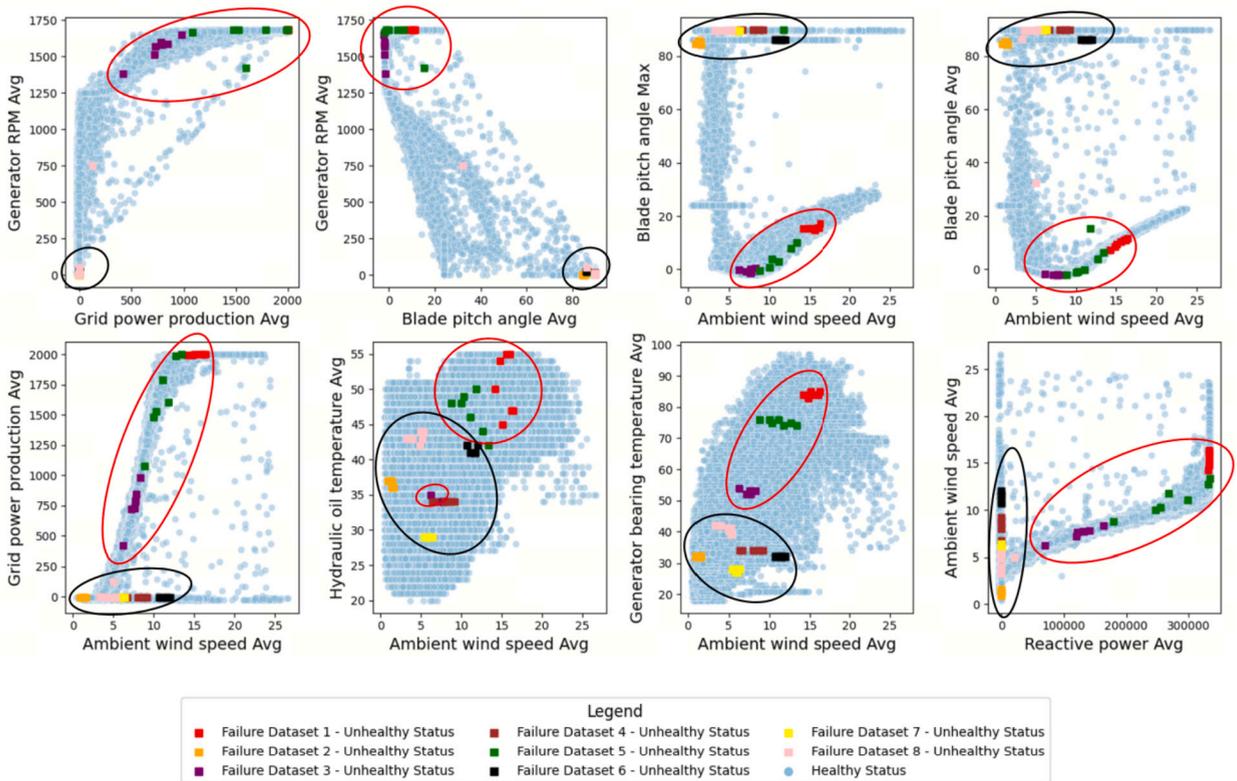

**Fig. 3.** Failure points correlation based on multi-parameter analysis.

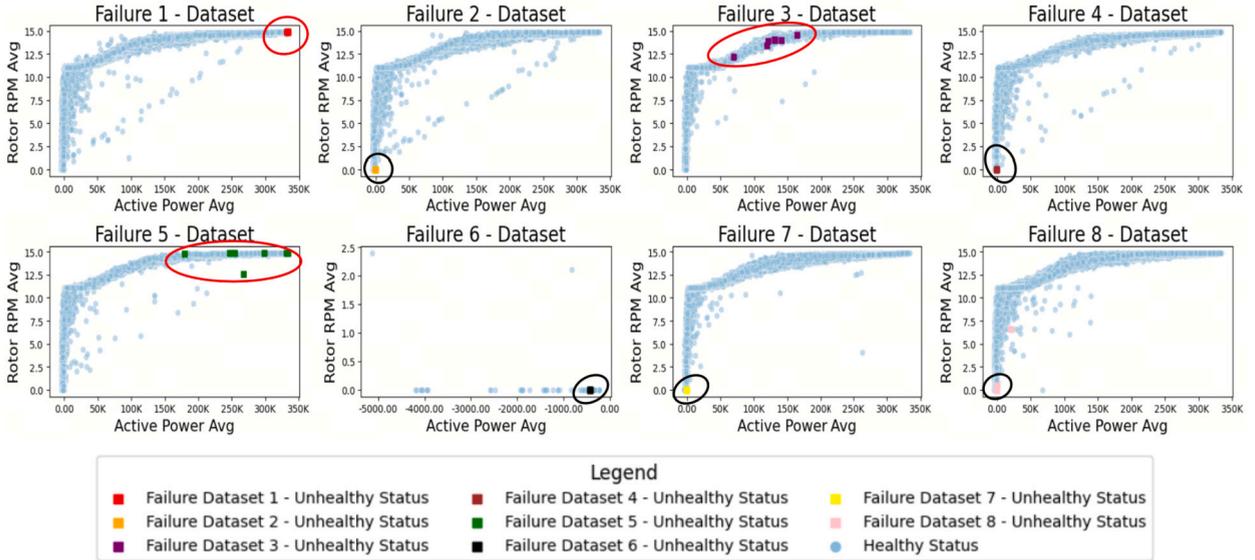

**Fig. 4.** Active power and rotor RPM analysis.

## 3. Experiment setting

### 3.1. Data preprocessing: general

**Time Series Data Preparation**: Time series format of the given WT's SCADA data can be equated as Equation (1) as follows:

$$D_i = \{d_1, d_2, \ldots, d_N\} \quad (1)$$

where $N$ is the total number of data logs available for that specific failure and could vary depending on the failure dataset, whereas $i$ represents the specific failure dataset and can range from 1 to 8, with the exception of 6.

Each data point $d_u$ contains $M$ parameters; Equation (2) can further concave the structuring for a given data point:

$$d_u = \{p_{u,1}, p_{u,2}, \ldots, p_{u,M}\} \qquad (2)$$

where the value of $M$ is taken to be 82, referring to the total number of parameters available in the SCADA dataset, and $u$ ranges from 1 to $N$.

**Linear Degradation Metric** ($L_i$): The current standards to evaluate RUL are only good at guessing the health status, failure probability, or curve-fitting of one specific parameter that is indicative of the health status in WTs [41–43], failing to provide a significantly longer time window. Since we are to forecast the RUL in advance, the standard methodologies are impractical. A study by Muneer et al. [44] employed a linear degradation model for future RUL estimation of NASA turbofan engines. This concept allows for the estimation of RUL by following a linearized trajectory of health degradation over time. We introduce this linear degradation metric to a forecasting step (detailed in Section 3.3) to formulate the functional basis for our $RUL_f$ methodology, i.e., forecasting. The linearization Equation (3) can further provide comprehension of this concept.

$$L_i = \{N-1, N-2, \ldots, 1, 0\} \qquad (3)$$

where $L_i$ is the RUL for the $i$-th dataset.

**Normalization**: Min-Max scaling is applied to bring all parameters to the same scale. From equation (2), each parameter, $p_j$ was taken for $N$ entries, as shown in equation (4), using minimum and maximum values for this parameter ($p_j$), a scaled value is obtained using equation (5).

$$p_j = \{p_{1,j}, p_{2,j}, \ldots, p_{N,j}\} \qquad (4)$$

$$p_{u,j}^{\text{scaled}} = \frac{p_{u,j} - \min(p_j)}{\max(p_j) - \min(p_j)} \qquad (5)$$

where $p_{u,j}^{\text{scaled}}$ is the scaled value of $p_{u,j}$.

**Input and Output Data**: The datasets are labelled into input ($X_i$) containing the SCADA data parameter and output (RUL, $Y_i$) data, delineated in Equation (6) & (7) respectively.

Input Data:

$$\overline{X_i} = \{\overline{X_{i,1}}, \overline{X_{i,2}}, \ldots, \overline{X_{i,N}}\} \qquad (6)$$

Output Data (*RUL*):

$$\overline{Y_i} = \{RUL_{i,1}, RUL_{i,2}, \ldots, RUL_{i,N}\} \qquad (7)$$

where $\overline{X_{i,u}}$ represents the 10-minute frequency log of $i$-th dataset at $u$-th index.

Following the preprocessing stage, the application of the SW approach facilitates the organization of data in a manner that allows for the representation of long-term disparities.

### 3.2. Data preprocessing: sliding window

Previous research on using WT SCADA data for DNNs has predominantly used 1D-DNNs or the sliding window (SW) technique for 2D input processing. The utilization of SW offers the benefit of effectively capturing temporal dependencies [45], and as stated previously, RUL is the measure for WT health's progressive degradation over time, therefore, the SW is employed to capture these time-varying dependencies in WT SCADA data. Putting consecutive data inputs in rows, with each output value (i.e., $RUL_u$, see equation (12)) linked to a window, creates a two-dimensional input data structure with time as the second dimension. The dimensions of the SW, specifically the width ($SW_{Width}$) and length ($SW_{Length}$), are contingent upon the number of rows (the number of parameters taken, i.e., $M = 82$) and the length of each individual row (the number of time logs taken, which is to be $l = 24$), respectively. The size of the SW ($SW_{Size}$) can be mathematically represented as equation (8). The stride is configured to be one unit, which implies the SW will move one log step at a time, creating a 2D input structure. The SW and the stride can be adjusted accordingly, tuning the ability to capture the time-varying dependencies. It is important to mention here that for the conversion of this 2D structured data into 3D, a depth of $SW_{Depth} = 1$ unit is applied.

$$SW_{Size} = SW_{Length} \times SW_{Width} \qquad (8)$$

In case of 3D DNN architecture, equation (8) is modified into equation (9):

$$SW_{Size} = SW_{Length} \times SW_{Width} \times SW_{Depth} \qquad (9)$$

After the implementation of SW, the input and output datasets are formulated as follows:

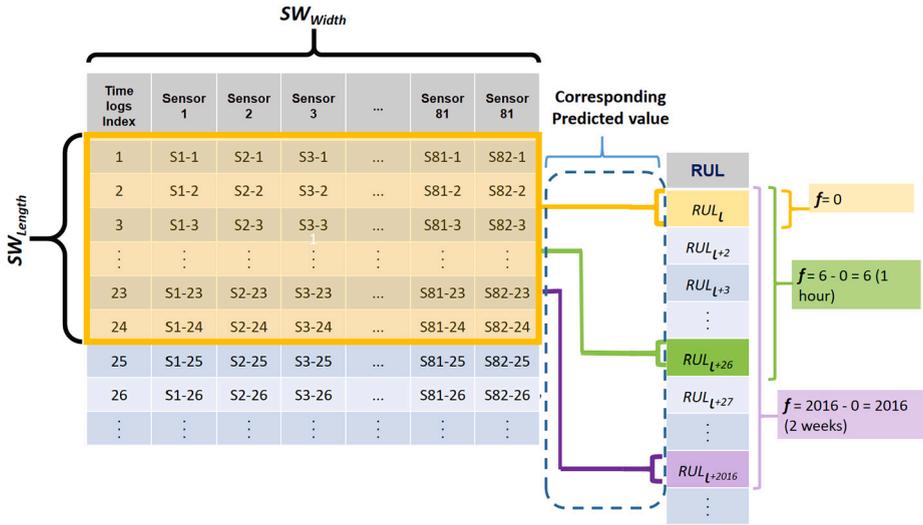

**Fig. 5.** The Sliding Window (SW).

Input data for dataset $i$:

$$X_{\text{input}, i} = \{X_1, X_2, \ldots, X_{N-l}\}_i \tag{10}$$

For a given input unit $X_q$ ($q$ ranging from 1 to $N - l$) represents the 2D—equal to $SW_{Size}$—at index $q$, and

$$X_{q,i} = \{d_u, d_{u+1}, \ldots, d_{u+l}\}_i \quad \text{where} \quad u \leq N - l \tag{11}$$

Output Data:

$$Y_{\text{output}, i} = \{RUL_l, RUL_{l+1}, \ldots, RUL_N\}_i \tag{12}$$

$RUL_w$ ($w$ ranging from $l$ to $N$) represents the output ($RUL_w$) at index $w$.

The exclusion of the final few data elements in the input dataset ($X_{input}$), ranging from index $N$-$l$+1 to index $N$, is a deliberate action taken to ensure the lengths of the input and output data remain the same. Following the same analogy for the output dataset ($Y_{output}$), the first data entry starts from index $l$, implying that the first $l$-1 number of outputs ($Y_1$ to $Y_{l-1}$) were excluded.

Following SW technique implementation, each input unit ($X_{q,i}$) carries a total of $l \times 10$ minutes (i.e., 240 minutes) worth of SCADA data. This enables a single time window containing multiple SCADA data logs to be linked to a single $RUL_w$ entry. The application of the sliding window technique restructured the input data into a 2D format, rendering it incompatible with 1D-DNN operations. However, this restructuring makes the data suitable for more advanced 2D and 3D DNN models, enabling more powerful analyses and predictions. The SW concept is further illustrated in Fig. 5.

To simplify the indices correspondence relation between the input and output, Equations (10) and (12) are revised as follows:

$$X'_{\text{input, i}} = \{X'_1, X'_2, \ldots, X'_R\}_i \tag{13}$$
$$Y'_{\text{output, i}} = \{RUL'_1, RUL'_2, \ldots, RUL'_R\}_i \tag{14}$$

where $R = N - l$. Both datasets, $X'_{input}$ and $Y'_{output}$, have an established correspondence with each other, i.e., $X'_z \to Y'_z$, where z defines indices for input and output datasets and varies from 1 to $R$. This association allows for the input dataset to be used for the prediction of the current health status, i.e., real-time RUL. The next step is to manipulate this established association in accordance with the RUL forecasting objective (future failure prediction), detailed in the subsequent section.

### 3.3. Data preprocessing: forecasting window

This research work introduces the novel concept of forecasting window (FW) on WT SCADA data for the WT RUL forecasting; FW determines how far in the future RUL or failure prediction can be done accurately. As mentioned in Section 3.2, each input data entry following the deployment of SW, is linked to a corresponding RUL value. For instance, the input log in $X'_{input}$ at index $z$ corresponds to a RUL value in $Y'_{output}$ at index $z$. The task of the FW is to establish a correspondence between the input value $X'_z$ at index $z$ and the RUL value, which is some predetermined steps ($f$, FW size) ahead at index $z+f$, i.e., $Y'_{z+f}$. Following the implementation of FW, the equations (13) and (14) are modified as follows:

$$X_{f, i} = \{X'_1, X'_2, \ldots, X'_{R-f}\}_i \tag{15}$$

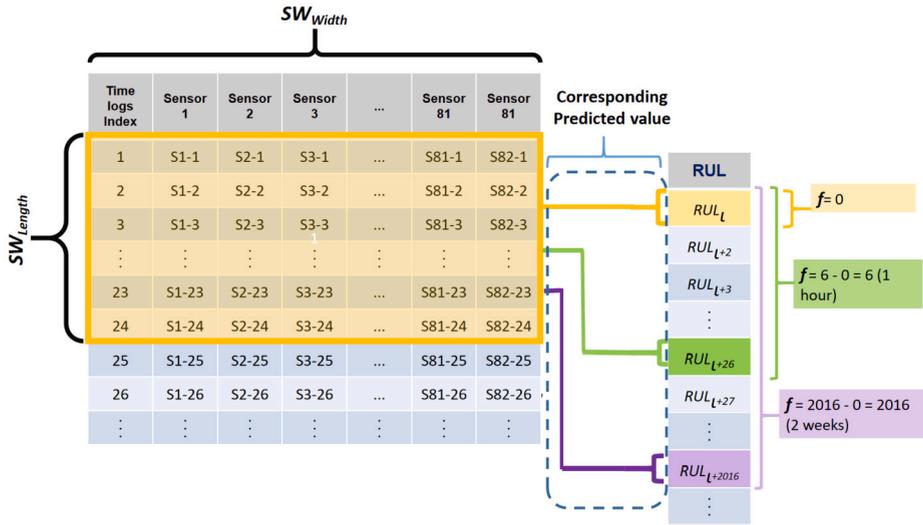

**Fig. 6.** The Forecasting Window (FW).

$$Y_{f, i} = \{RUL'_f, RUL'_{f+1}, \ldots, RUL'_R\}_i \tag{16}$$

Both of the above equations (15) & (16) can be rewritten to simplify the indices as follows:

$$X_{F, i} = \{X_{F1}, X_{F2}, \ldots, X_{F,g}\}_i \tag{17}$$

$$Y_{F, i} = \{RUL_{F1}, RUL_{F2}, \ldots, RUL_{F,g}\}_i \tag{18}$$

The simplified concept from equation (10) to (18) is as follows:

$$\begin{bmatrix} X_{N-l} \to RUL_N \\ X_{N-l-f} \to RUL_N \\ \text{(renaming for the sake of convenience, i.e.,} \\ X_{N-l-f} = X_{F,g} \quad \& \quad RUL_N = RUL_{F,g}) \\ X_{F,g} \to RUL_{F,g} \end{bmatrix}_i$$

This can be visualized in Fig. 6.

### 3.4. Model architecture

A cohort of research has utilized different DNN models to analyze WT SCADA data. Lei et al. (2019) [46] demonstrated the effectiveness of long short-term memory (LSTM) model in processing and classifying time-series data, showing outstanding performance in long-term time-series datasets. Zhang et al. (2023) [24] developed an adaptive multivariate time-series convolutional network (AdaMTCN) to extract enriched features from SCADA data. Qin et al. (2023) [47] enhanced pitch system fault diagnosis using a Multi-Channel Attention Long Short-Time Memory Network (MCA-LSTM), which uses multi-channel AM to detect pitch bearing anomalies and hub faults with remarkable accuracy. Xiang et al. [48] had introduced a convolutional neural network (CNN) with an integrated AM for fault detection, enhancing the model's ability to issue early warnings and predict turbine failures before they escalate into more significant issues. However, since our objective is to forecast RUL over much longer period of 2-weeks while bypassing feature engineering, a much more sophisticated model architecture is required. In this context, our study proposes two new hybrid DNN models, ForeNet-2d and ForeNet-3d, which combines the strengths of CNNs, LSTMs, and AM, enhancing their ability to identify and prioritize relevant features, leading to more accurate and timely fault predictions.

#### 3.4.1. ForeNet-2d

Xiang et al. [48] took advantage of AM [49]—to manipulate training weights—in combination with CNN and LSTM to give early WT failure alerts. However, for our application, since the RUL metric reflects the linear degradation of the WT health over the entire period of its operational life, which can be few months, for such long period forecasting, a custom designed model is required which specialized towards catching these long term dependencies. Therefore, taking insights from Xiang et al., this research study proposes its own custom-optimized ForeNet-2d DL model architecture for the purpose of $RUL_f$. The designed model aims to efficiently capture intricate temporal patterns and dependencies present in multivariate time series data.

The combination of CNN and LSTM models has proven effective across various domains, particularly for time-series tasks [50,51]. These studies highlight that CNNs excel in feature extraction, while LSTMs are adept at retaining temporal patterns [52]. The synergy of these models creates a robust tool for time-series predictions. Additionally, the incorporation of AM in various applications has further enhanced model efficiency [53,54]. The effectiveness of such hybrid structures has been consistently validated. The ForeNet-2d model leverages this powerful combination through a sequential architecture composed of CNN, LSTM, and AM layers, each dedicated to a specific function. The CNN block, consisting of three convolutional layers, is designed to extract temporal discrepancies across a wide array of parameters, enabling the model to map data features to complex patterns via convolutional operations. This allows for feature extraction and regression from 2D signal data while avoiding the complexities associated with multi-dimensional samples [48]. However, given the time-series nature of the task, where long-term temporal patterns are crucial, a single LSTM layer is introduced to bolster the model's ability to remember these patterns [51]. Following the LSTM layer, an AM is applied to emphasize the parameters most indicative of turbine failure. This layered structure harnesses the strengths of each individual model component, delivering superior forecasting capabilities compared to any single model.

**Table 3**
ForeNet-2d Model Structure.

| Layer (type) | Output Shape | Param # |
| --- | --- | --- |
| InputLayer | (None, 24, 82) | 0 |
| Conv1D | (None, 22, 64) | 15,808 |
| Conv1D | (None, 20, 64) | 12,352 |
| Conv1D | (None, 18, 128) | 24,704 |
| LSTM | (None, 18, 64) | 49,408 |
| Reshape | (None, 18, 64) | 0 |
| Attention | (None, 18, 64) | 0 |
| Flatten | (None, 1152) | 0 |
| Dense | (None, 1) | 1,153 |

**Table 4**
ForeNet-3d Model Structure.

| Layer (type) | Output Shape | Param # |
| --- | --- | --- |
| InputLayer | (None, 24, 82, 1) | 0 |
| Conv2D | (None, 22, 80, 64) | 640 |
| Conv2D | (None, 20, 78, 32) | 18,464 |
| Conv2D | (None, 20, 78, 1) | 33 |
| Attention | (None, 20, 78, 1) | 0 |
| Multiply | (None, 20, 78, 32) | 0 |
| Multiply | (None, 20, 78, 32) | 0 |
| Multiply | (None, 20, 78, 32) | 0 |
| Flatten | (None, 49,920) | 0 |
| Dense | (None, 1) | 49,921 |

The model layout is as follows (see Table 3 and Fig. 7): The model's input layer is designed to handle multivariate time series data with a shape of ($l = 24$, $M = 82$). The CNN component commences with a configuration consisting of two Conv1D layers (from the Keras machine learning algorithm); Conv1D not to be confused with the 2D input shape, which was achieved with the application of SW (Section 3.2). The two Conv1D layers are equipped with 64 filters and a kernel size of 3. The feature extraction process is carried out by the convolutional layers, which involve sliding over the time series data and applying rectified linear unit (ReLU) activation to introduce non-linearity. This non-linearity is crucial in addressing the vanishing gradient problem, allowing the network to learn complex patterns and hierarchical representations effectively. The subsequent Conv1D layer is composed of 128 filters, which serve to further enhance the feature representation. Thus, the input data goes through a total of three Conv1D layers before the LSTM layer, ensuring that the model enables efficient feature extraction, hierarchical representation learning, and translation invariance, enhancing the model's ability to capture temporal patterns for accurate $RUL_f$.

In order to effectively capture temporal dependencies, specifically the RUL linear degradation behaviour, over the entire tenure of the failure dataset, the LSTM layer is positioned subsequent to the CNN block. The LSTM network is composed of 64 units and is set up to output the sequences of hidden states by enabling the return sequences parameter to be true. This enables the LSTM model to effectively transmit pertinent temporal information across consecutive time steps.

Subsequently, the AM is employed to selectively emphasize training weights manipulation, thereby enhancing the predictive accuracy of the model. The use of AM in this DNN model is necessitated by the lack of feature engineering step. The output of the attention layer is subsequently flattened for further processing. The flattened attention output is connected to a Dense layer, which is responsible for converting the extracted representations into the output used for forecasting RUL predictions. The Dense layer is configured with a single unit, and the final prediction of the RUL is obtained by applying a linear activation function.

It is to point out here that Xiang et al. [48] used AM to manipulate training weights in order to obtain faster convolution. In contrast, this research utilized AM to assess the influence of each parameter on a specific failure to manipulate the training weights and circumvent the constraints of feature engineering.

### 3.4.2. ForeNet-3d

To the best of our knowledge, the application of a 3D DNN model architecture for the processing of time series predictions using WT SCADA data has not been previously explored; however, some fault classification applications have been implemented [24]. Also, 3D-DNNs have been effectively used to capture temporal dependencies in other research fields from the pseudo-3D data structure derived from 2D input data (pseudo-3D-DNN) [55]. 3D DL architecture has a key advantage over 2D since spatial learning can be achieved in three dimensions simultaneously. Therefore, this study attempts a novel 3D DNN architecture, outlined in Table 4, to process WT SCADA data for time series application, i.e., $RUL_f$, optimized towards WT RUL forecasting. It is important to note that, unlike the ForeNet-2d model, the ForeNet-3d model does not include an LSTM component. Instead, it incorporates an element-wise multiplication section designed to enhance the discrepancies captured during the CNN convolution process. The exclusion of the LSTM layer in ForeNet-3d is due to its distinct architectural design compared to ForeNet-2d, as well as the observed performance results during testing, which indicated that the revised structure sufficiently captured the necessary temporal patterns without the need for an LSTM layer. It should be noted that even though various studies (as referenced previously) indicate a performance boost in CNN through the incorporation of LSTM, it is usually for a layered architecture (as in ForeNet-2d), which can be accounted for by following a different architecture design, such as ForeNet-3d's parallel-bypass multiplication architecture.

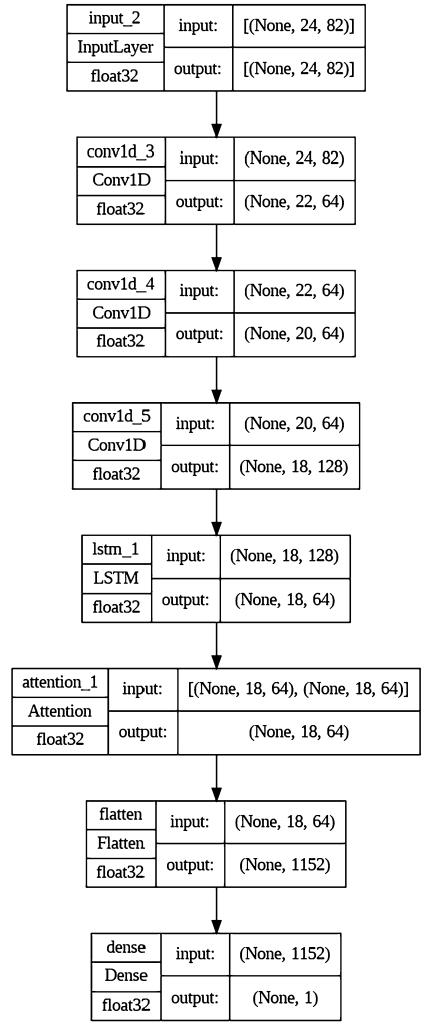

**Fig. 7.** ForeNet-2d Architecture.

One of the primary tasks of utilizing such an architecture was to convert the input data structure for DNN into a 3D format. The formulation of the 2D input shape for the ForeNet-2d has been previously discussed in Section 3.2, and to transform this given 2D input shape into a 3D shape, the depth of unit 1 ($SW_{Depth}$) is utilized; see Equation (9).

The DNN architecture begins with the utilisation of two consecutive Conv2D layers. Each of these layers is configured with 64 filters and a kernel size of 3. The purpose of these convolutional layers is to extract significant features from the input data and introduce non-linearity by utilising the ReLU activation function. Subsequently, a Conv2D layer is employed, utilising a kernel size of 1. The purpose of this layer is to function as a softmax activation layer, which is responsible for extracting further discrepancies while avoiding further reductions in the data shape. The AM is employed to highlight important temporal events by highlighting weights, allowing the model to concentrate on crucial points within the time series data. The model integrates element-wise multiplication between the output of the second Conv2D layer and the attention weights. The objective of this multiplication process is to give priority to the discrepancies captured during the CNN section of the model and, at the same time, retain the key contributing features contributing towards a specific failure. The process of element-wise multiplication is repeated two additional times, resulting in multiple iterations that enhance significant temporal patterns while diminishing less informative features. After the multiplication layers, the model proceeds to flatten the data in order to facilitate the transition to a Dense layer. The final output of the WT's $RU\ L_f$ prediction is represented by a Dense layer with a single unit and a linear activation function. Each neuron in the Dense layer is connected to every neuron in the previous layer.

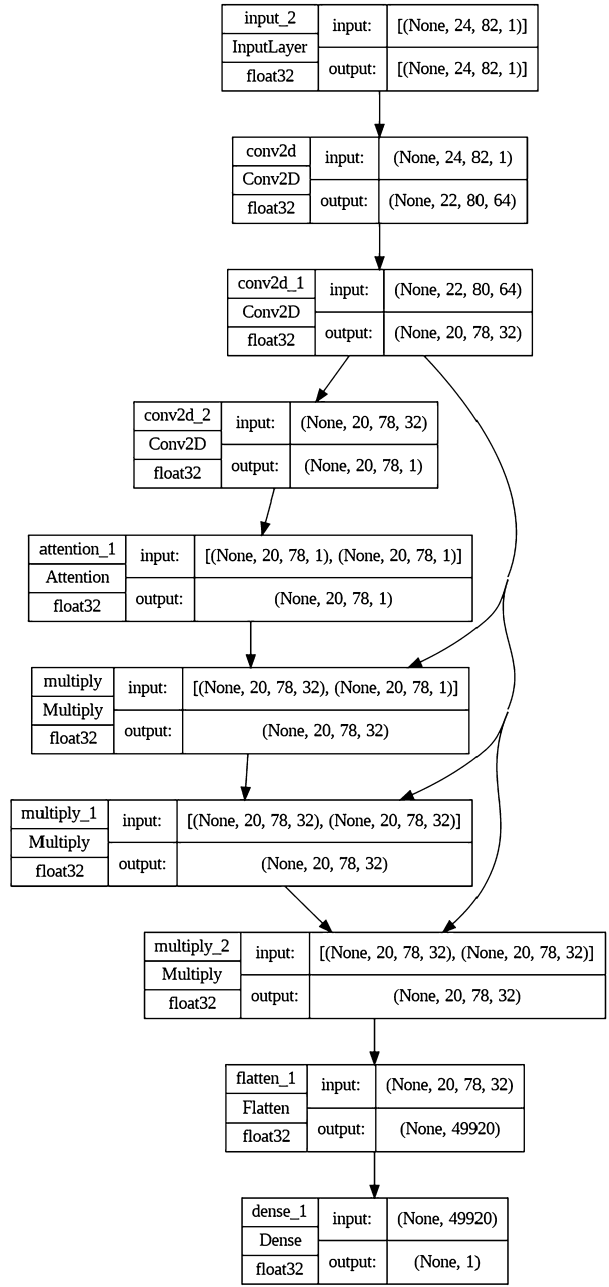

**Fig. 8.** ForeNet-3d Architecture.

The incorporation of AM in the model's design highlights the influential elements that affect the model's output and is used due to the same reason for by-passing feature engineering steps as mentioned in Section 3.4.1. However, it can be hypothesised that the influence of AM is relatively diminished in comparison to the ForeNet-2d, as, in contrast to the integration of the AM in a sequential fashion—a linear series architecture—in the ForeNet-2d (Fig. 7), the ForeNet-3d incorporates the AM in a parallel-bypass arrangement, as indicated in Fig. 8. The model architecture consists of a total of 102,593 trainable parameters. The complete model architecture summary can be seen in Table 4.

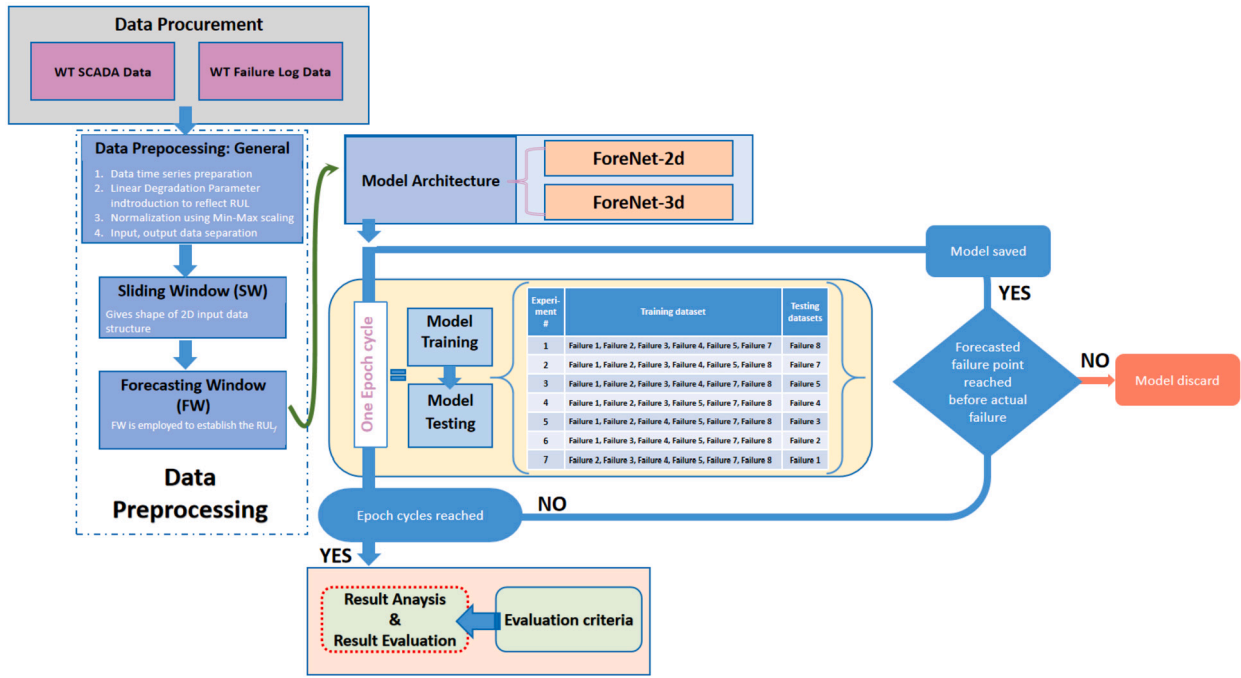

**Fig. 9.** Experimental setting overview.

### 3.5. Evaluation criteria

The evaluation criteria for $RUL_f$ evaluates the accuracy in future failure prediction with reference to its actual occurrence timing. This can be quantified by accessing the temporal disparity ($D_k$), equated in Equation (19), between the projected end-of-life cycle (failure) for the WT and the observed failure time. For example, in the event that the model's prediction deviates by 5 hours from the observed time of failure, the model's forecasting accuracy is deemed to be inaccurate by a margin of 5 hours.

$$D_k = R\hat{U}L_k - RUL_k \tag{19}$$

where $R\hat{U}L_k$ is the forecasted RUL at time $k$.

### 3.6. Model training and testing

The experimental setup was devised such that for every instance of failure, both models were trained using the remaining failure datasets and subsequently evaluated for that particular failure based on the criteria specified in Section 3.5. As summarized in Section 3.5, the model's performance evaluation is based on the accurate failure forecast, which corresponds to a zero value at the end of the failure dataset, representing WT breakdown. However, it was theorized that the model with a higher level of effectiveness based on the root mean square error (RMSE) parameter would naturally demonstrate superior ability to achieve better curve-fitting, resulting in increased accuracy in forecasting progression towards zero value, indicating failure. To this end, the training process of the models was carefully executed with a specific focus on reducing the loss function, particularly RMSE with help of Adam optimizer [56,57]. The model training process was purposefully designed to systematically reach the prognosticated failure points before the actual failure happened. This tactical approach involved permitting the ongoing trained model to be saved only in cases where the forecasted failure threshold was reached preemptively by testing the trained model after each epoch via a looped training mechanism, reducing the likelihood of post-forecast failures. A total of 14 experiments were conducted, i.e., 7 RUL forecast predictions for all seven datasets for both models. Each experiment utilized the remaining failure datasets to train the model, consequently resulting in varied dataset combinations and epochs needed to attain a particular degree of accuracy for each experiment setting. In order to maintain a consistent experimental framework, 10 epochs were instantiated, and the one with the highest model accuracy was saved for forecasting applications. It is imperative to acknowledge that a plausible augmentation in the model's forecasting capabilities could be achieved via an increase in epoch count, thereby suggesting a prospective avenue for further refinement in forecasting performance.

The overall experiment summary can be visually encapsulated in Fig. 9, providing a visual representation of the key facets of the conducted experiments.

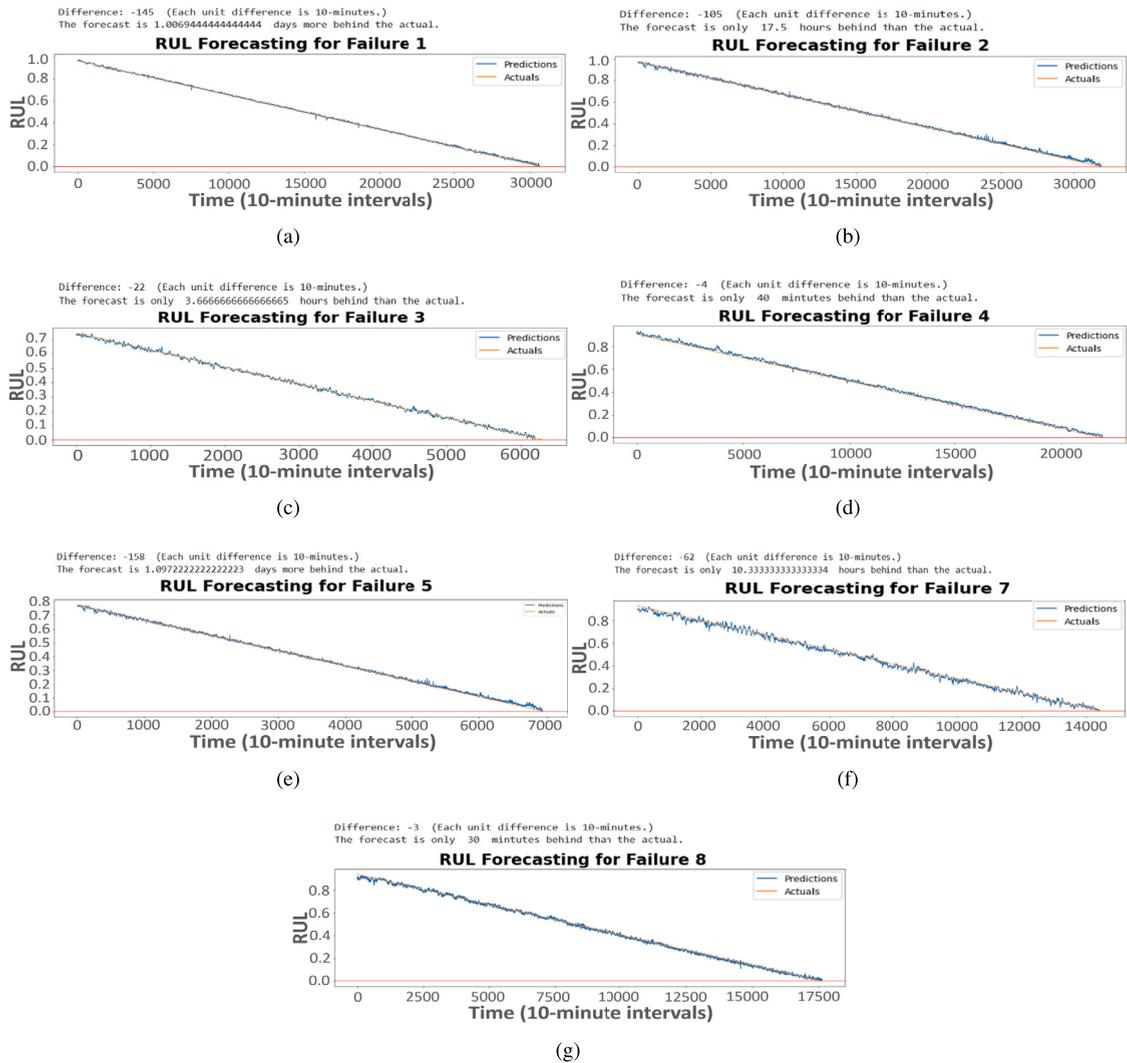

**Fig. 10.** ForeNet-2d Results: The figure shows a two-week $RUL_f$. The X-axis shows turbine life (SCADA data log: logs are recorded with a 10-minute difference) and the Y-axis shows RUL degradation.

**Table 5**
Experiment Result Detail: The table mentions the detail of model deviation from actual failure.

| Model | Experiment Result Detail (2-weeks $RUL_f$) |
| --- | --- |
| **ForeNet-2d** | |
| Failure 1 | 1 day behind the actual failure |
| Failure 2 | 17.5 hours behind the actual failure |
| Failure 3 | 3.7 hours behind the actual failure |
| Failure 4 | 40 minutes behind the actual failure |
| Failure 5 | 1.1 days behind the actual failure |
| Failure 7 | 10.3 hours behind the actual failure |
| Failure 8 | 30 minutes behind the actual failure |
| **ForeNet-3d** | |
| Failure 1 | 1.8 days behind the actual failure |
| Failure 2 | 2.6 hours behind the actual failure |
| Failure 3 | 10 minutes behind the actual failure |
| Failure 4 | 9 hours behind the actual failure |
| Failure 5 | 1.2 hours behind the actual failure |
| Failure 7 | 3 hours behind the actual failure |
| Failure 8 | 6.2 hours behind the actual failure |

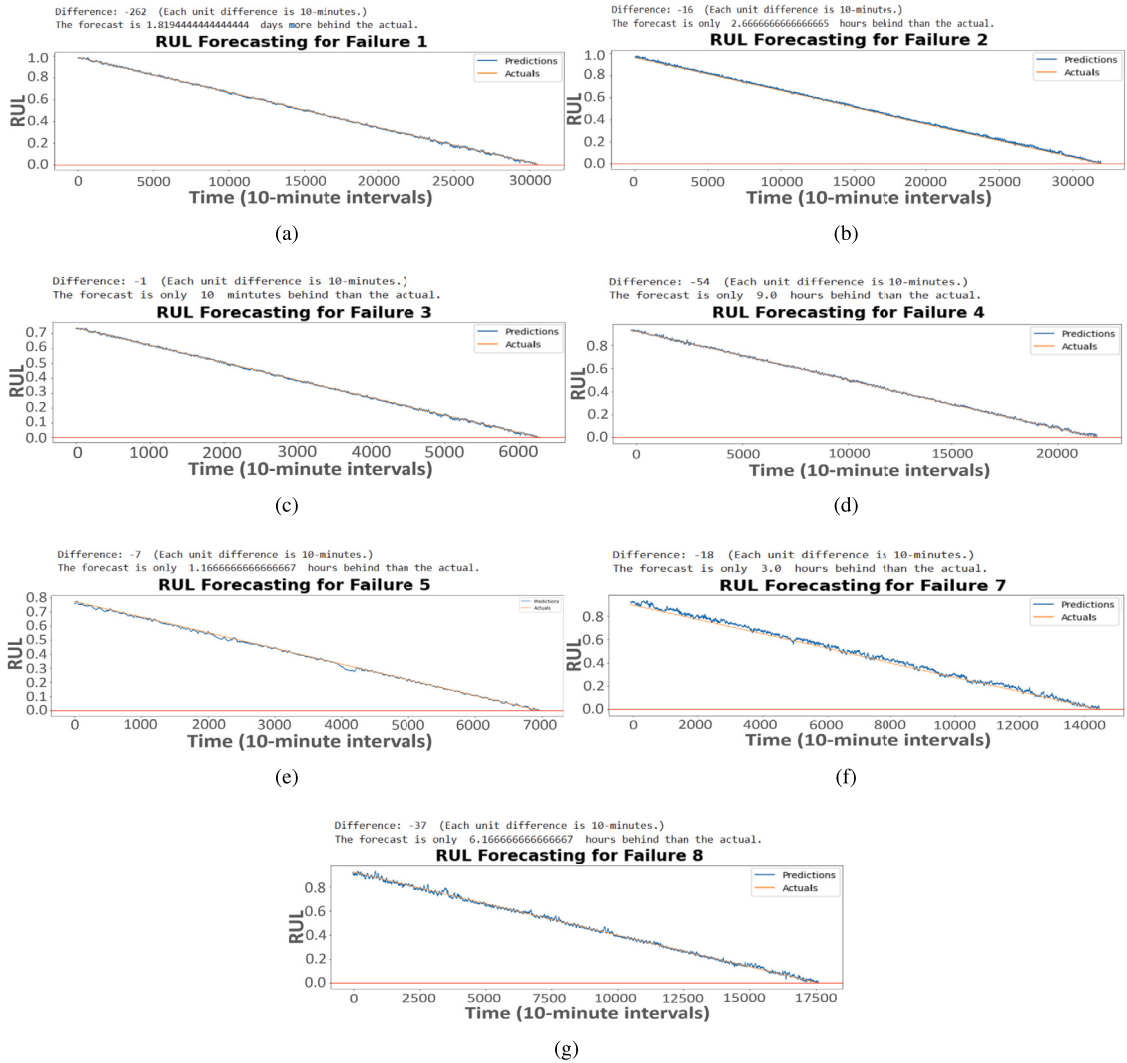

**Fig. 11.** ForeNet-3d Results: The figure shows two-week $RUL_f$. The X-axis shows turbine life (SCADA data log: logs are recorded with 10 minutes difference) and the Y-axis RUL degradation.

## 4. Results and analysis

The testing process for both models is presented in Figs. 10 and 11; Fig. 10 consist ForeNet-2d testing excerpt for Failure 1 (Fig. 10a), Failure 2 (Fig. 10b), Failure 3 (Fig. 10c), Failure 4 (Fig. 10d), Failure 5 (Fig. 10e), Failure 7 (Fig. 10f) and Failure 8 (Fig. 10g), similarly, in case of ForeNet-3d, Fig. 10 showcase Failure 1 (Fig. 11a), Failure 2 (Fig. 11b), Failure 3 (Fig. 11c), Failure 4 (Fig. 11d), Failure 5 (Fig. 11e), Failure 7 (Fig. 11f) and Failure 8 (Fig. 11g). Moreover, the results from all 14 experiments are summarized in Tables 5 & 6. The ForeNet-2d demonstrated superior performance for Failures 4 and 8, exhibiting a minimal discrepancy of 40 and 30 minutes, respectively. In contrast, the outcomes with the least precision were observed for Failures 1 and 5, exhibiting a deviation of slightly more than a day each. In the case of ForeNet-3d, optimal performance was exhibited for Failure 3, demonstrating a marginal forecast difference of only 10 minutes, while Failure 7 being a bit more modest around 10 hours. Conversely, the least precise outcome was observed for Failure 1 of 1.8 days. Here, the model accuracy parameter can be further elaborated by taking an example from one of the experiments, i.e., Failure 4 from the ForeNet-2d experiments. The projected failure forecast from this experiment is 40 minutes before the actual failure. This shows that the model was able to forecast RUL two weeks in advance, which was off by only 40 minutes.

This research underscores the significance of minimizing deviations between the DNN forecast and the actual failure instance, the crucial significance of which is ensuring a sufficient and reliable time window for advance maintenance scheduling leading to PdM. As a result, the model's capacity to predict beyond the 2-week mark is limited due to increased inaccuracy post-2-week FW, posing a constraint on the DNN's predictive capabilities. From Table 6 & 5, the least accurate results obtained are:

**Table 6**
RUL forecast accuracy comparison—The negative sign shows that the forecasted failure happened before the actual failure.

|  | ForeNet-2d | ForeNet-3d |
| --- | --- | --- |
|  | $D_k$ | $D_k$ |
| Failure 1 | -145 | -262 |
| Failure 2 | -105 | -16 |
| Failure 3 | -22 | -1 |
| Failure 4 | -4 | -54 |
| Failure 5 | -158 | -7 |
| Failure 7 | -62 | -18 |
| Failure 8 | -3 | -37 |

**Table 7**
The table presents a comparison of 2 weeks $RUL_f$ accuracy ($D_k$) in comparison to failure location and data-logs available.

| Failure | ForeNet-2d | ForeNet-3d | Data-logs Available | Failure Location |
| --- | --- | --- | --- | --- |
|  | Difference ($D_k$) |  |  |  |
| 1 | -145 | -262 | 31,779 | Transformer |
| 2 | -105 | -16 | 33,005 | Hydraulic group |
| 3 | -22 | -1 | 8,480 | Gearbox |
| 4 | -4 | -54 | 24,101 | Hydraulic group |
| 5 | -158 | -7 | 9,011 | Generator |
| 7 | -62 | -18 | 16,661 | Hydraulic group |
| 8 | -3 | -37 | 19,814 | Hydraulic group |

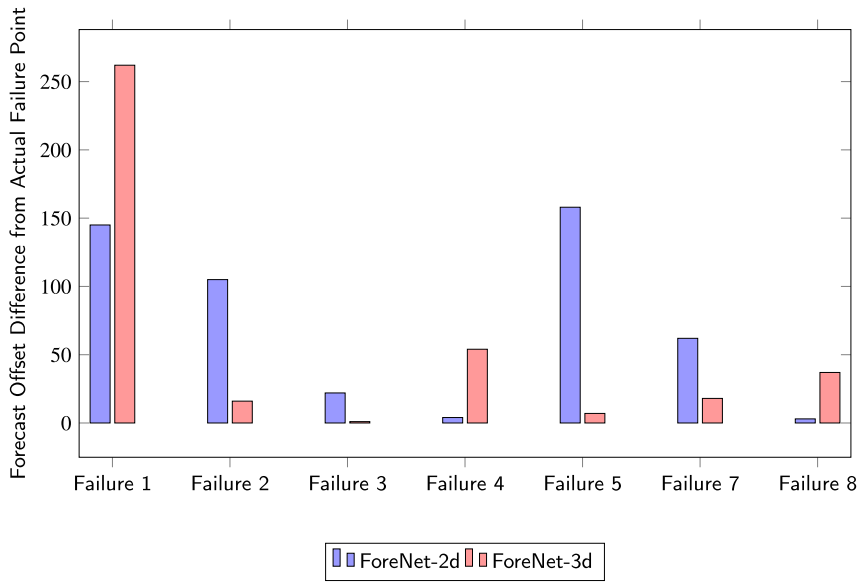

**Fig. 12.** Cluster Graph: Comparison of Models Results for FW of 2 weeks.

1. Failure 5 for ForeNet-2d → 1.1 days difference
2. Failure 2 for ForeNet-3d → 1.8 days difference

Rounding off to 2 days to compensate for the model inaccuracies, we are left with a 12-day reliable maintenance window. This temporal buffer remains notably advantageous for the orchestration of preemptive PdM measures. It is to emphasize here that providing the models with additional training on akin datasets holds the potential to widen the FW.

A comparison between the both ForeNet models is drawn in Fig. 12; a simplistic overview of this comparison reveals that both models exhibit lower forecast capabilities for Failure 1, while, for the remaining failure cases, both models capabilities differ comparatively. It can be inferred that the ForeNet-2d exhibited superior performance compared to the ForeNet-3d for failures 1, 4, and

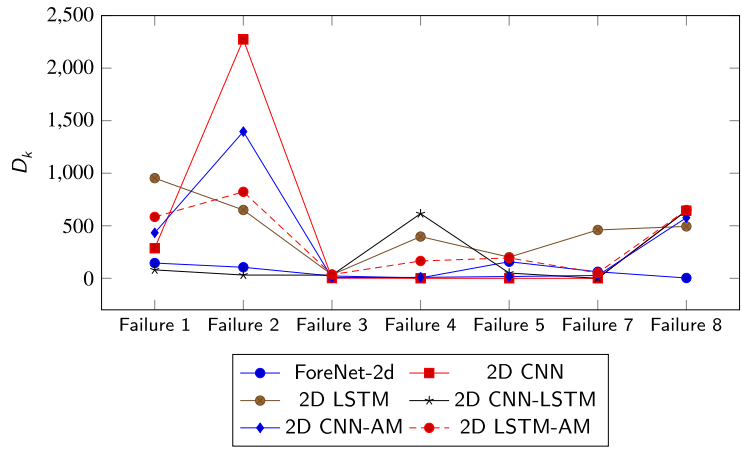

**Fig. 13.** $RUL_f$ result comparison for 2D models.

8. Conversely, the ForeNet-3d demonstrated better performance for the remaining failure cases. Also, a general visual overview of the graph tells us that the ForeNet-3d showcased marginally superior performance overall compared to the ForeNet-2d.

Table 7 presents the results of the $RUL_f$ accuracy in relation to different failure types and failure lengths (i.e., $N$). The negative sign in the "Difference" column indicates that the forecasted failure precedes the actual failure occurrence; ensuring this alignment is crucial in the context of $RUL_f$ as the goal is to proactively anticipate and forecast WT breakdowns prior to their actual occurrence. The column denoted "Data Logs Available" in the table shows the duration before each failure instance, representing a temporal window ultimately leading towards breakdown at the last log entry. The table also specifies the location of the failure for each instance, such as the transformer, hydraulic group, gearbox, and generator. Both models accurately predict the RUL for each failure incidence, regardless of location. The models showed varying forecast accuracy ($D_k$) across experiments, with variances in $RUL_f$ ranging from a few logs to several hundred. In the event of Failure 1, specifying transformer failure, the ForeNet-2d demonstrated a deviation of -145 log values, whereas for the ForeNet-3d, this was -262 logs. Conversely, in the case of Failure 4, related to the hydraulic group, the respective $D_k$ values were -4 and -54. The reason for this disparity between these failure point forecasts may be attributed to the differing number of failure instances associated with the hydraulic group and the transformer.

As mentioned in Table 7, there are four failure points related to the hydraulic group, providing ample training data for this particular failure type, whereas only one failure instance, Failure 1, is available for the transformer. Besides this, the close correlation exhibited by the failure instances associated with the hydraulic group, as shown in Fig. 3, is testimony to smaller $D_k$, resulting in a better forecast. In the case of the Failure 3, affiliated with the gearbox, both models exhibited notable performance, with deviations of only -22 and -1 log values, respectively. It is worth noting that only one instance of gearbox failure was available for analysis. Thus, such close proximity might take into account the same correlation, discussed prior in Section 2.3 with reference to Fig. 3 and 4.

Since Section 3.4 asserted the superiority of the ForeNet model architecture over other models [24,46–48], it is indispensable to justify this assertion by comparing ForeNet models with these predecessors. Therefore, Figs. 13 and 14 illustrate the preeminence of ForeNet-2d and ForeNet-3d relative to other models. This superiority is largely attributed to their advanced architectures and the integration of the AM. Moreover, the importance of AM is clearly demonstrated in Figs. 15 and 16, which present the average $D_k$ for the 2D and 3D models. Fig. 15 shows that ForeNet-2d significantly outperforms the CNN-LSTM model—essentially ForeNet-2d without AM. Similarly, CNN-AM and LSTM-AM surpass their counterpart models, CNN and LSTM, respectively. This pattern is also evident in Fig. 16, where ForeNet-3d outperforms CNN-M (which has the same architecture as ForeNet-3d but lacks AM) and other prominent models (GoogLeNet, AlexNet, and ResNet-50, all of which lack AM). Thus, it is reasonable to conclude that AM plays a critical role in enhancing the accuracy of the ForeNet models, particularly because the $RUL_f$ methodology and the models circumvent standard feature engineering; without AM, it is quite challenging for the models to effectively weight the key parameters contributing to failure progression.

## 5. Conclusion

With the employment of the FW concept, this study was successfully able to use DL to provide a 2-week lead time window for PdM in hopes of mitigating the O&M cost in WT. Considering the most accurate forecasted RUL was off by only 10 minutes of the actual RUL and the majority were off by only a few hours, while the least accurate deviated by 1.8 days, the majority were off by only a few hours. The proposed models were able to employ multi-parametric SCADA data to base adaptive RUL forecasts for different failure types (based on location) by making use of the existing coherence among the apparent post-failure parametric behaviours. Since all of the available parameters were used for this application and the impact of each failure for a certain failure type was evaluated using the model architecture, the need for feature engineering was successfully bypassed, and thus any chance of human error during this step. However, it must be acknowledged that the $RUL_f$ methodology did not account for categorizing the failure type within its two-week advance forecast window, which could have been more desirable. Unfortunately, achieving this level of specificity would require additional datasets specific to each component-based failure to train the models adequately, enabling the identification of the exact component where the failure is likely to occur. In spite of that, in the case of $RUL_f$ for WT as a whole, both models exhibited much superior performance in comparison to previous models adopted by the research cohort due to their optimization towards extracting linear health degradation from a multi-parametric dataset due to careful integration of AM which allowed for identifying failure-specific parameters and enhanced extraction towards temporal discrepancies. It is important to acknowledge that the design of the ForeNet-3d exclusively utilized a CNN architecture, with AM just slightly enhancing the training weights. Whereas the ForeNet-2d architecture was based on a combination of CNN and LSTM, AM was employed as a much more integral part of the model here to manipulate training weights. Augmentation to the quantity of training datasets pertaining to particular failure types could potentially lead to enhanced accuracy in forecasting RUL for these failures.

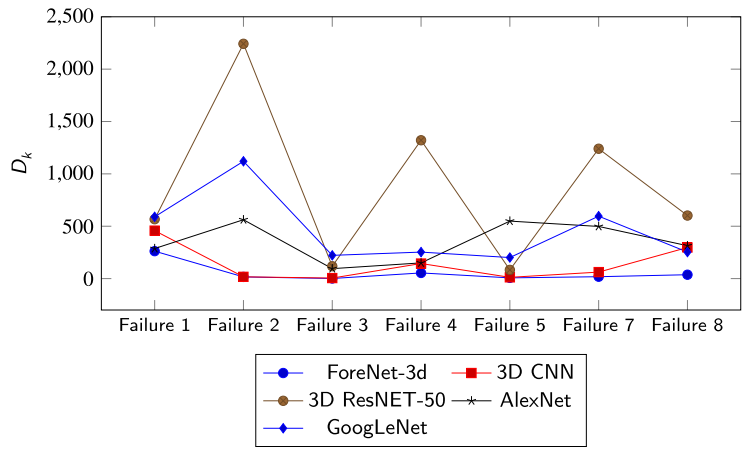

**Fig. 14.** $RUL_f$ result comparison for 3D models.

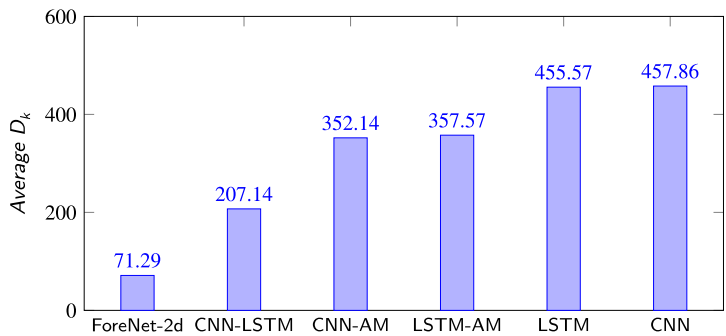

**Fig. 15.** Average $D_k$ comparison for 2D models in ascending order.

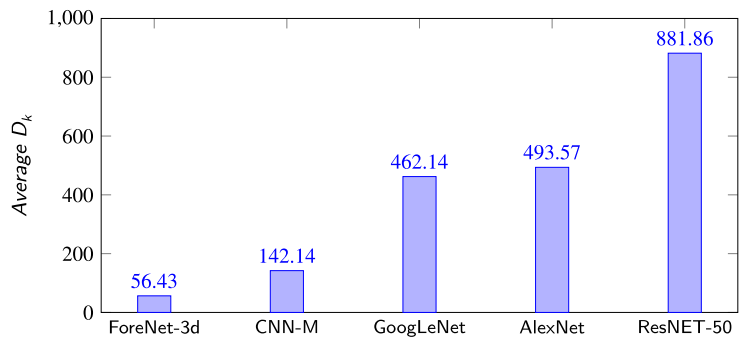

**Fig. 16.** Average $D_k$ comparison for 3D models in ascending order. *Note: CNN-M is based on the ForeNet-3d architecture with the absence of AM.*

## 6. Future work

For future work, it is plausible that $RUL_f$ methodology could also potentially facilitate the categorization of failure locations or failure components, contingent upon the availability of sufficient training data pertaining to a specific failure. Furthermore, it is important to evaluate the model's performance on significantly larger datasets that encompass a greater number of failures. While this research work was able to capture the failure discrepancies over a longer duration and forecast RUL, it is also possible to utilize shorter time windows (i.e., FW) to proactively identify alarm warnings a few minutes to hours in advance. It would also be interesting to investigate how the accuracy of the models is affected by varying the number of parameters used or by making an incremental increase in the number of parameters while assessing the effect on accuracy. But this falls beyond the scope of this research study. As this research study establishes the use of 3D DNN for WTs' SCADA data, it would be interesting to access the performance of $RUL_f$ while deploying some other famous 3D DNN architectures. The two models proposed in this study, although optimised for $RUL_f$, might potentially be helpful in accessing other key features also using WT SCADA data.

**CRediT authorship contribution statement**

**Syed Shazaib Shah:** Writing – review & editing, Writing – original draft, Visualization, Validation, Software, Resources, Methodology, Investigation, Formal analysis, Data curation, Conceptualization. **Tan Daoliang:** Writing – review & editing, Validation, Supervision, Resources, Methodology, Conceptualization. **Sah Chandan Kumar:** Validation, Software, Formal analysis.

**Funding**

This work was supported by Beihang University Science Center for Gas Turbine Project under Grant 2022-B-V-004 and Chinese Government Scholarship under Grant 2021SLJ008269.

**Declaration of competing interest**

The authors declare that they have no known competing financial interests or personal relationships that could have appeared to influence the work reported in this paper.

**Data availability**

Data will be made available on request.